\documentclass[final]{IEEEtran}

\usepackage{booktabs}
\usepackage{cite}
\usepackage[cmex10]{amsmath}
\usepackage{amssymb}

\usepackage{pgfplots}
\definecolor{lines-1}{RGB}{228,26,28}
\definecolor{lines-2}{RGB}{55,126,184}
\definecolor{lines-3}{RGB}{77,175,74}
\definecolor{lines-4}{RGB}{152,78,163}
\definecolor{lines-5}{RGB}{255,127,0}
\definecolor{lines-6}{RGB}{255,255,51}
\definecolor{lines-7}{RGB}{166,86,40}
\definecolor{lines-8}{RGB}{247,129,191}
\definecolor{lines-9}{RGB}{153,153,153}
\pgfplotscreateplotcyclelist{myCycleList}{
	color=lines-1,semithick,mark=o,mark size=2\\%
	color=lines-2,semithick,mark=star,mark size=2\\%
	color=lines-3,semithick,mark=square,mark size=2\\%
	color=lines-4,semithick,mark=diamond,mark size=2\\%
	color=lines-5,semithick,mark=triangle,mark size=2\\%
	color=lines-8,semithick,mark=Mercedes star,mark size=2\\%
	color=lines-7,semithick,mark=|,mark size=2\\%
	color=lines-9,semithick,mark=oplus star,mark size=2\\%
}
\pgfplotsset{
	compat=1.14,
	width =\columnwidth, 
	height=.8\columnwidth,
	ylabel absolute, ylabel style={yshift=-0.2cm},
	xlabel absolute, xlabel style={yshift=0.2cm},
	label style={font=\scriptsize},
	tick label style={font=\scriptsize},
	legend style={font=\scriptsize,cells={align=left}},
	grid=both,
	minor grid style={dotted},
}

\usetikzlibrary{spy}

\newcommand{\define}{\triangleq}

\newcommand{\transpose}{\intercal}

\begin{document}



\title{Achievable Information Rates for Nonlinear Fiber Communication via End-to-end Autoencoder Learning}%



\author{
 Shen Li, Christian H\"ager,
    Nil Garcia, Henk Wymeersch
    
\thanks{The authors are with  the Department of Electrical Engineering, Chalmers University of Technology, Gothenburg, Sweden. Christian H\"ager is also with the Department of Electrical and Computer Engineering, Duke University, Durham, USA. email: shenl@student.chalmers.se. 
This work is part of a project that has received funding from the
European Union's Horizon 2020 research and innovation programme under
the Marie Sk\l{}odowska-Curie grant agreement No.~749798. }    
}

\maketitle                  



  \begin{abstract}
    Machine learning is used to compute achievable information rates (AIRs) for a simplified fiber channel. The approach jointly optimizes the input distribution (constellation shaping) and the auxiliary channel distribution to compute AIRs without explicit channel knowledge in an end-to-end fashion.
  \end{abstract}


\section{Introduction}




Fiber transmission rates can be increased by multi-level quadrature amplitude modulation (M-QAM) formats, which require higher input power and are thus more susceptible to nonlinear impairments such as  nonlinear signal-noise interaction (NLSNI). Conventional techniques to deal with NLSNI include improved detector designs\cite{Ho2005, Lau2007, Tan2011} and optimized modulation formats\cite{Lau2007,Haeger2013,ref5}. The achievable transmission rates are themselves upper-bounded by the channel capacity, which is unknown for optical channels with NLSNI, even for simplified nondispersive scenarios, though upper\cite{Keykhasravi2017} and lower\cite{Turitsyn2003, Keykhasravi2017, Yousefi2011} capacity bounds have been established.  


A different approach for constellation or detector design is to rely on machine learning and deep learning, including \cite{Oshea2017, Zibar2016, Doerner2018, Haeger2018, Haeger2018isit, Karanov2018, Lee2018}. Recently, autoencoders (AE) have emerged as a promising tool for end-to-end design  and have been shown to lead to good performance for wireless \cite{Oshea2017, Doerner2018}, noncoherent optical\cite{Karanov2018}, as well as visible light communication\cite{Lee2018}. 


%
In this paper, we develop an AE for a simplified memoryless fiber channel model. 
It is shown that the AE approach can be used to establish tight lower bounds on the channel capacity by computing  achievable information rates (AIR)\cite{Arnold2006, Djordjevic2005, Secondini2013, Fehenberger2015}. Moreover, the AE can approach maximum likelihood (ML) performance and leads to optimized constellations that are more robust against NLSNI than conventional QAM formats. 




\section{Simplified Fiber Channel Model}

Similar to\cite{Turitsyn2003, Ho2005, Lau2007, Tan2011, Yousefi2011, Haeger2013, Keykhasravi2017}, we consider a simplified memoryless channel for fiber-optic communication which is obtained from the nonlinear Schr\"odinger equation by neglecting dispersion. The resulting per-sample model is defined by the recursion 
\vspace{-3mm}
\begin{align}
    {x}_{k+1}={x}_{k}e^{j L\gamma \left | {x}_{k} \right |^{2}/K}+{n}_{k+1}, \quad 0\le k <K, \label{eq:FOchannel}
\end{align}
where $x_0 = x$ is the (complex-valued) channel input, $y = x_K$ is the channel output, ${n}_{k+1} \sim \mathcal{C N }({0},P_{N}/K)$, $L$ is the total link length,  $P_{N}$ is the noise power, and $\gamma$ is the nonlinearity parameter. The model assumes ideal distributed amplification and $K \to \infty$. The channel input $x$ is drawn randomly from an $M$-point constellation with $\mathbb{E}\{|{X}|^2\}=P_{\text{in}}$, where $P_{\text{in}}$ is the input power.  


Even though dispersive effects are ignored, the model still captures some of the nonlinear effects encountered during realistic transmission over optical fiber, in particular nonlinear phase noise (NLPN). 
The main interest in this channel model lies in the fact that the channel probability density function (PDF) $p(y|x)$ is known analytically\cite{Turitsyn2003, Ho2005, Yousefi2011}. This allows us to compare the AE performance to an ML detector and benchmark the obtained AIRs using known capacity bounds.  

\section{Proposed Autoencoder Structure}

\begin{figure*}
   \centering
        \includegraphics{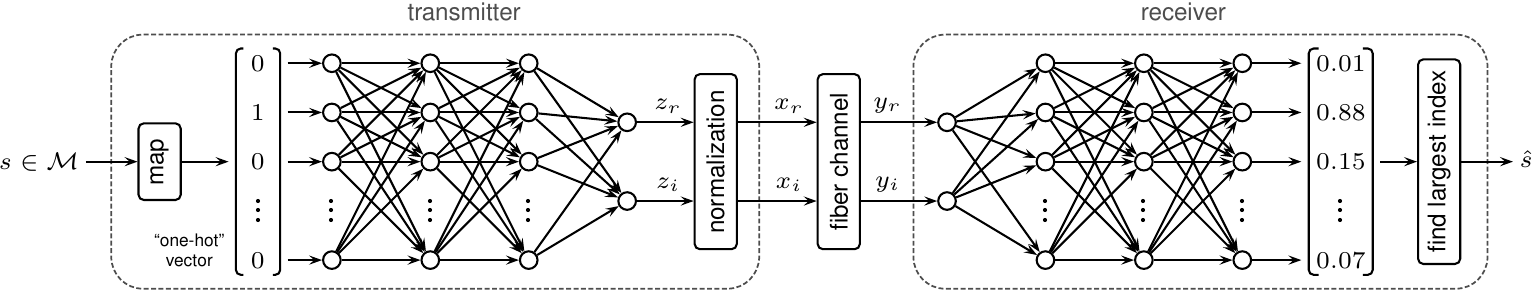}
    \caption{Autoencoder structure assuming $2$ hidden layers in both the transmitter and receiver neural network.}
    \label{fig:figure1}
\end{figure*}

In machine learning, an AE is a neural network (NN) which consists of two parts: an encoder maps an input $s$ (e.g., an image) to a lower-dimensional representation or code and a decoder attempts to reconstruct the input from the code. It has recently been proposed to interpret all components of a communication system, consisting of a transmitter, channel, and receiver, as an AE\cite{Oshea2017}. This allows for end-to-end learning of good transmitter and receiver structures. 

The AE structure used in this paper is shown in Fig.~\ref{fig:figure1} and will be described in the following. 
The goal is to transmit a message $s$ chosen from a set of $M$ possible messages $\{1,2,...,M\} \define \mathcal{M}$. Following \cite{Oshea2017}, the messages are first mapped to $M$-dimensional "one-hot" vectors where the $s$-th element is $1$ and all other elements are $0$. The one-hot vectors denoted by $\mathbf{u}$ are the inputs to a transmitter NN, which consists of multiple dense layers of neurons. Each neuron takes inputs from the previous layer and generates an output according to $z_{\text{out}}=f(\mathbf{w}^{\transpose}\mathbf{z}_{\text{in}} + b)$, where $\mathbf{w}$ is a vector of weights, $b \in \mathbb{R}$ is a bias, and $f(\cdot)$ is an activation function, here considered to be a sigmoid or tanh function. 
%
%
The values of the two transmitter output neurons ($z_r$ and $z_i$ in Fig.~\ref{fig:figure1}) are used to form the channel input. To meet the average power constraint, a normalization is applied using $M$ different training inputs to the NN. Then the normalized output is assumed to be sent over the channel, leading to an observation $y$. The real and imaginary parts of $y$ are taken as the input to a receiver NN, the output of which we denote by $f_y(s')\in [0,1], s' \in \mathcal{M}$, where we assume a sigmoid in the last layer and then normalize the sum of the output to 1. Finally, we set $\hat{s}=\arg\max_{s'}f_y(s')$. 


The AE is trained using many batches of training data averaging over different messages and channel noise configurations. In particular, the weights and biases of all neurons in both the transmitter and receiver NN are optimized with respect to $\frac{1}{N} \sum_{i=1}^{N} \ell(u^{(i)}_{s}, f_y(s')^{(i)})$, where
\begin{equation}
\ell(u^{(i)}_{s},f_y(s')^{(i)})=
-u^{(i)}_{s}\log f_y(s')^{(i)}.
\end{equation}
is the cross-entropy loss, $N$ is the batch size (a multiple of $M$), and the superscript refers to different training data realizations, the subscript $s$ refers to the $s^{th}$ element of $\mathbf{u}^{(i)}$. The optimization is performed using a variant of stochastic gradient descent with an appropriate learning rate. 





\section{Achievable Information Rates}



The AE can be used to determine lower bounds on the mutual information 
\begin{align} 
    \label{eq:MI}
    I(X,Y)=\sum_{x}\int p(x,y)\log_2\frac{p(y|x)}{p(y)}\mathrm{d}y
\end{align} 
as follows\cite{Arnold2006, Djordjevic2005, Secondini2013, Fehenberger2015}. We normalize $f_y(s')$ with respect to $s'$ and consider it as a distribution over $x$. Then,
 $f_y(x)p(y)$ is a valid joint distribution over $x$ and $y$, so that, due to the non-negativity of the Kullback-Leibler divergence,  $\mathrm{KL}(p(x,y)||p(y)f_{y}(x))\ge0$. Straightforward manipulations then yield
\begin{align}
    I(X,Y)\ge \sum_{x}\int p(x,y)\log\frac{f_y(x)}{p(x)}\mathrm{d}y, \label{eq:AIR}
\end{align}
which can easily be evaluated via Monte Carlo integration. The right-hand side of \eqref{eq:AIR} is the AIR of the AE. Both the mutual information and the AIR are lower bounds on the channel capacity. 

\section{Performance Analysis}

For the numerical results, we assume $L=5000\text{ km}$, $\gamma=1.27$, and $P_N = -21.3 \text{ dBm}$. The number of iterations to simulate the fiber model \eqref{eq:FOchannel} is set to $K=50$, which is sufficient to approximate the true asymptotic channel PDF\cite{Ho2005}. The AE is trained separately for different values of $P_{\text{in}}$ using the Adam optimizer in TensorFlow. The AE parameters are summarized in Tab.~\ref{tab:structure}. 


	
\begin{table}
	\centering
	\caption{Autoencoder parameters}
	\footnotesize
	\begin{tabular}{llll|lll}
		\toprule
		& \multicolumn{3}{c}{transmitter} & \multicolumn{3}{c}{receiver} \\
		\midrule
		\textbf{layer} &1 & 2--6 &7 & 1 & 2--7& 8 \\
		\textbf{neurons} &  $M$ &  $M$ & 2  &  2&  $M$  &  $M$  \\
		\textbf{$f(\cdot)$}& - &tanh & linear & - &tanh&sigm.\\
		\bottomrule
	\end{tabular}
	\label{tab:structure}%
\end{table}



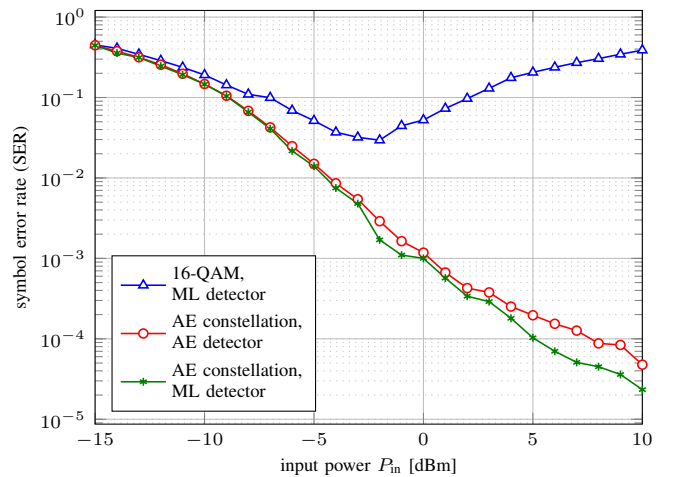
\begin{figure}
	\centering
	\begin{tikzpicture}
	\begin{semilogyaxis}[
		xmin=-15, xmax=10,
		xlabel={input power $P_{\text{in}}$ [\text{dBm}]},
		ylabel={symbol error rate (SER)},
		cycle list name=myCycleList,
		legend pos=south west,
		legend entries = {16-QAM{,} \\ML detector,AE constellation{,}\\AE detector,AE constellation{,}\\ML detector},
		legend cell align=left,
		ylabel style={yshift=-.05cm},
		width = \columnwidth,
	]
	\addplot+[color=blue, mark=triangle*, mark options={mark size=2.0pt, fill=white}] table[
		x=power,
		y=16QAM,
	] {./fig2data.txt};
	\addplot+[color=red, mark=*, mark options={mark size=1.7pt, fill=white}] table[
		x=power,
		y=AE,
	] {./fig2data.txt};
	\addplot+[color=black!50!green, mark=asterisk, mark size=1.5pt, mark options={solid}] table[
		x=power,
		y=AEML,
	] {./fig2data.txt};
	\end{semilogyaxis}
	\end{tikzpicture}
	\caption{SER as a function of $P_{\text{in}}$ for $M=16$.}
	\label{fig:SER}
\end{figure}

\begin{figure} 
\centering
\includegraphics[width=0.47\columnwidth]{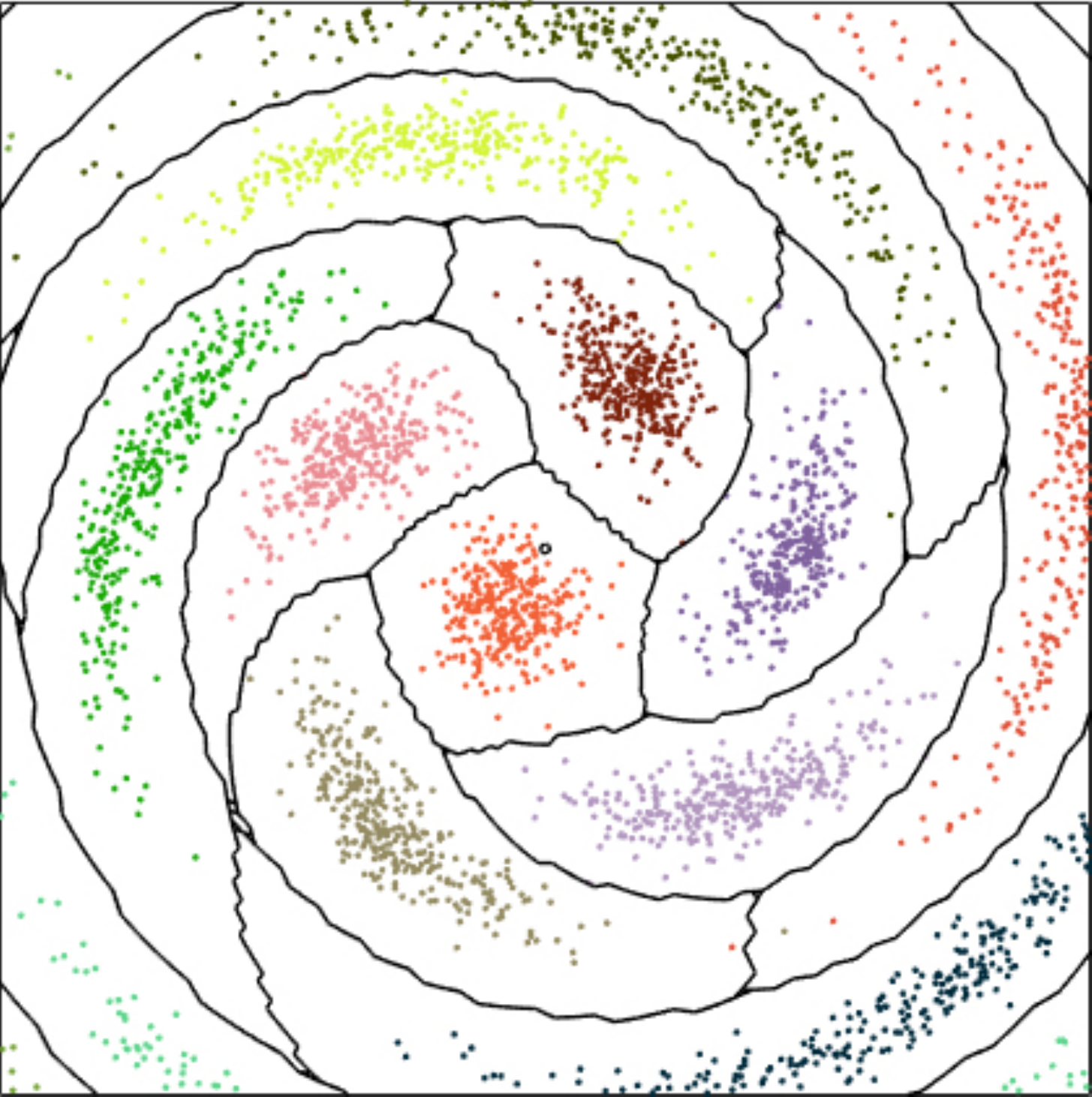}
\hfill
\includegraphics[width=0.47\columnwidth]{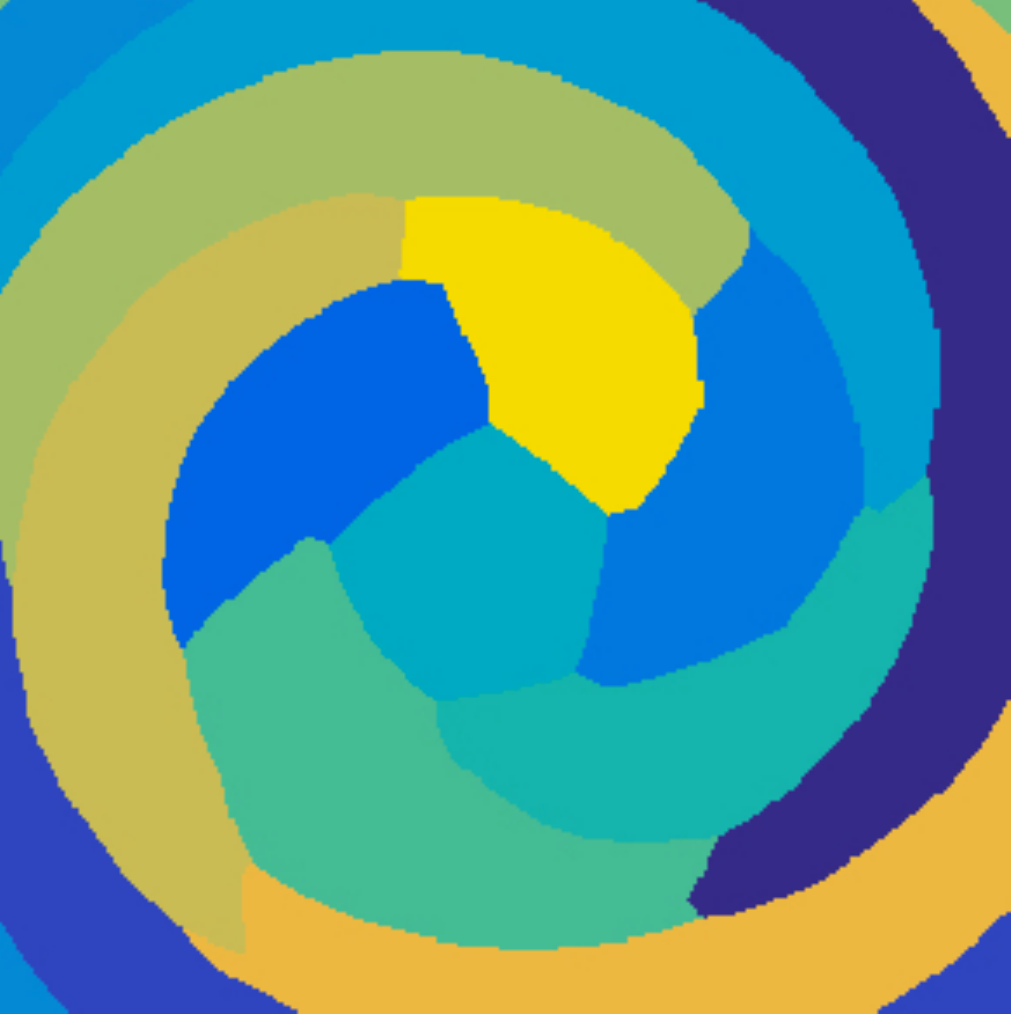}
\caption{ML decision boundaries for the AE constellation at $P_{\text{in}} = 0$ dBm (left) and learned AE decision regions (right).}
\label{fig:dr}
\end{figure}

We start by comparing the symbol error rate (SER), i.e., $p(s\neq \hat{s})$, of the AE to the SER of an ML detector applied to (a) standard 16-QAM and (b) the signal constellation optimized by the AE. The results are shown in Fig.~\ref{fig:SER}. The optimal input power for 16-QAM under ML detection is around $-2$ dBm, after which the SER increases due to NLPN. The SER of the AE decreases with input power, showing that the AE can find more suitable constellations in the presence of NLPN. If we replace the receiver part of the AE with an ML detector, the SER improves only slightly. This indicates that the AE can not only learn good constellations, but also learn to approximate the correct channel distribution, thus achieving near-ML performance. To visualize this, in Fig.~\ref{fig:dr}, we compare the effective decision regions implemented by the AE after training (right) to the optimal ML decision regions for the optimized AE constellation at $P_{\text{in}} = 0$ dBm (left), showing excellent agreement.


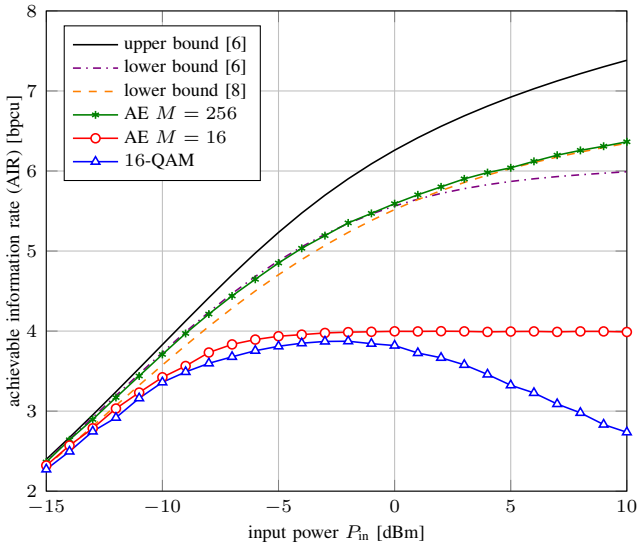
\begin{figure}
	\centering
	\begin{tikzpicture}
	\begin{axis}[
		xmin=-15, xmax=10,
 		ymin=2, ymax= 8,
 		y label style={at={(axis description cs:0.11,.5)}, anchor=south},
		xlabel={input power $P_{\text{in}}$ [\text{dBm}]},
		ylabel={achievable information rate (AIR) [bpcu]},
		cycle list name=myCycleList,
		legend pos=north west,
		width = 1.05\columnwidth,
		height = 0.9\columnwidth,
		legend cell align=left,
 		legend style={font=\scriptsize, row sep=-0.05cm},
	]
	\addplot+[color=black, no markers] table[
		x=P_in,
		y=upper,
	] {./fig4data.txt};
	\addlegendentry{upper bound\cite{Keykhasravi2017}}


 	\addplot+[no markers, dashdotted, color=violet] table[
 		x=P_in,
 		y=lower2,
 	] {./fig4data.txt};
	\addlegendentry{lower bound\cite{Keykhasravi2017}}
	
 	\addplot+[no markers, dashed, color=orange] table[
 		x=P_in,
 		y=lower1,
 	] {./fig4data.txt};
	\addlegendentry{lower bound\cite{Yousefi2011}}

	\addplot+[color=green!50!black, mark=asterisk, mark size=1.5pt, mark options={solid, green!50!black}] table[
		x=P_in,
		y=Iae256,
	] {./fig4data.txt};
	\addlegendentry{AE $M = 256$}
	
	\addplot+[color=red, mark=*, mark options={mark size=1.7pt, fill=white}] table[
		x=P_in,
		y=Iae16,
	] {./fig4data.txt};
	\addlegendentry{AE $M = 16$}
	
	\addplot+[color=blue, mark=triangle*, mark options={mark size=1.9pt, fill=white}] table[
		x=P_in,
		y=I16qamML,
	] {./fig4data.txt};
	\addlegendentry{$16$-QAM}

	\end{axis}
	\end{tikzpicture}
	\caption{Comparison of the AIR of the AE to various information-theoretic capacity bounds and $16$-QAM.}
	\label{fig:AIR}
\end{figure}

In Fig.~\ref{fig:AIR}, the AIR of the AE for $M = 16$ and $M = 256$ is shown. We first  compare the case $M=16$ to the mutual information $I(X;Y)$ assuming $16$-QAM as the input distribution. Note that the mutual information \eqref{eq:MI} can also be evaluated via Monte Carlo integration since the channel PDF $p(y|x)$ is known. As expected, the mutual information for $16$-QAM decreases with input power, whereas the AIR of the AE flattens out at the maximum value $\log_2 16 = 4$. Lastly, we compare the AIR of the AE for $M=256$ to three information-theoretic bounds on the channel capacity: the solid black line corresponds to a recently derived upper bound\cite{Keykhasravi2017}, whereas the dashed and dash-dotted lines correspond to lower bounds based on a  Gaussian\cite{Keykhasravi2017} and half-Gaussian\cite{Yousefi2011} input distribution, respectively. The AIR of the AE closely follows the maximum of the two lower bounds, slightly exceeding them at the crossover point at around $0$ dBm. These results indicate that the optimized AE constellations are close to being capacity-achieving and that the upper capacity bound can be further tightened.  



\section{Conclusions}

We have presented an autoencoder approach to communicating over a simplified nonlinear fiber channel. The approach allows for end-to-end learning of good signal constellations and the  channel posterior distribution. It was shown that the autoencoder can learn constellations that are robust to nonlinear phase noise and outperform conventional $M$-QAM constellations. Moreover, near-ML performance can be obtained without explicit channel knowledge. We also evaluated the achievable information rate of the AE, showing that the obtained lower capacity bounds are comparable to, and sometimes slightly exceed, two existing lower bounds for the considered nonlinear fiber channel model.



\bibliographystyle{IEEEtran}






\end{document}